\documentclass[aps,prb,twocolumn,groupedaddress,showpacs,superscriptaddress]{revtex4-1}
\usepackage{graphicx,amsmath}
\usepackage{amssymb}
\usepackage{xcolor}

\begin{document}
\title{Enhanced Quantum behavior on frustrated Ising model: Quantum Approximate Optimization Algorithm study}
\author{Seunghan Lee}
\affiliation{Department of Liberal Studies, Kangwon National University, Samcheok, 25913, Republic of Korea}
\author{Hunpyo Lee}
\email{Email: hplee@kangwon.ac.kr}
\affiliation{Department of Liberal Studies, Kangwon National University, Samcheok, 25913, Republic of Korea}
\affiliation{Quantum Sub Inc., Samcheok, 25913, Republic of Korea}
\date{\today}

\begin{abstract}

We investigated the quantum effects of a frustrated Ising model on a two-dimensional square lattice using the Quantum Approximate Optimization Algorithm (QAOA). While strong spin frustration is known to induce quantum fluctuations at low temperatures, previous classical approaches—restricted to binary (up/down) spin configurations—have been insufficient to fully capture the quantum contributions of frustration. In this study, we introduced a quantitative metric to evaluate the quantum effects arising from frustration and employed QAOA to differentiate between classical and quantum regimes. Notably, we found that in the weakly frustrated region, QAOA measurements rarely capture first excited states, as they are energetically well separated from the ground state. In contrast, near the quantum phase transition point, excited states appear more frequently in QAOA measurements, highlighting the increased role of quantum fluctuations.
\end{abstract}

\maketitle

\section{Introduction\label{Introduction}}

The frustrated Ising model, which undergoes a quantum phase transition and represents a class of combinatorial optimization problems, serves as an effective testbed for evaluating the performance of newly developed computational methods. A key advantage of this model is that the energy gap between the ground state and the first excited state can be systematically tuned, making it ideal for benchmarking algorithmic efficiency and accuracy. The energy gap is a critical parameter that directly influences the complexity and precision of combinatorial optimization tasks; when the gap becomes small—particularly near the quantum phase transition—the problem becomes significantly more difficult to solve accurately. To address this challenge, a variety of numerical techniques have been developed, including Monte Carlo simulations, machine learning algorithms, simulated annealing, and quantum annealing~\cite{Jin2012, Jin2013, Park2022, Lee2024, Romer2025}.

Recent advances in quantum hardware—including gate-based quantum computers and quantum annealers have significantly enhanced their ability to address combinatorial optimization problems encountered in fields such as economics, computational science, and machine learning~\cite{Preskill2018, Bharti2022, arute2019, postler2022, nation2025, Johnson2011, Higgott2019, King2021, Park2024(1), Park2024(2)}. These quantum machines are believed to offer new insights into the underlying quantum nature of such problems by exploiting quantum entanglement, which emerges during the evolution of quantum states associated with optimization tasks. Consequently, it is of great interest to investigate how quantum behavior depends on the energy gap, using quantum algorithms.

The Quantum Approximate Optimization Algorithm (QAOA) stands out as a prominent method for finding approximate solutions to combinatorial optimization problems using quantum entanglement~\cite{farhi2014, farhi2015, farhi2019, choi2019, farhi2020, zhou2020, moussa2022, blekos2024}. Inspired by the Quantum Adiabatic Algorithm (QAA), it employs Trotterization to simulate time-dependent adiabatic evolution and approximate the ground state of a system, thereby addressing the original optimization task~\cite{farhi2000, farhi2001}. Unlike QAA, which performs optimization through continuous adiabatic evolution between an initial and a final Hamiltonian, QAOA adopts a discrete, layered structure that alternates between cost and mixer Hamiltonians. When a combinatorial optimization problem is formulated in terms of binary variables, its objective function can be encoded into a cost Hamiltonian within the QAOA framework. The alternating application of cost and mixer Hamiltonians forms a parameterized quantum ansatz, which serves as a sequence of unitary time-evolution operators. A classical optimizer iteratively updates the variational parameters based on measurement outcomes from the quantum ansatz, aiming to approximate the optimal solution. The number of layers $p$ determines the ansatz depth; as $p$ increases, the ansatz explores a larger portion of the solution space and can yield better approximations. In the limit $p = \infty$, QAOA converges to quantum adiabatic computing, which theoretically guarantees the optimal solution but at the cost of significantly longer run times.

In this paper, we focused on how the energy gap between the ground state and the first excited state influences both the accuracy and quantum behavior on the QAOA approach. To explore this, we considered the frustrated Ising model on a two-dimensional $4 \times 4$ square lattice with nearest-neighbor feromagnetic (FM) interaction $J_1$ ($J_1>0$) and diagonal antiferomagnetic (AF) interaction $J_2$ ($J_2<0$). This model, under periodic boundary conditions, exhibits a quantum phase transition between FM and stripe AF phases at zero temperature, where the energy gap $\Delta/J_1$ between the ground state and the first excited state vanishes at $\vert J_2 \vert/J_1=0.5$. Notably, the ratio $\vert J_2 \vert/J_1$ provides a systematic means of controlling $\Delta/J_1$. Using the QAOA approach with an appropriate number of layers, via computing the order parameters of spin configurations, we first confirmed the FM-to-stripe AF phase transition at $\vert J_2 \vert/J_1=0.5$, which is well-known from previous Monte Carlo, simulated annealing and D-Wave quantum annealing studies~\cite{}. Next, assuming that the exact ground state and first excited state energies are known, we computed the probability of obtaining the correct solution as a function of the number of layers. We confirmed that a larger $\Delta/J_1$ allows for a high success probability even with fewer layers thank to absence of quantum effects, whereas a smaller $\Delta/J_1$ requires deeper ansatz to achieve similar accuracy. Finally, we observed that the expectation value of the energy converges to the ground state energy in the large $\Delta/J_1$, while it deviates in $\Delta/J_1$ due to enhanced quantum fluctuations.

The remainder of this paper is organized as follows. Section~\ref{Model} provides a detailed description of the frustrated Ising model and the QAOA simulation. In Section~\ref{Result}, we present and analyze the results. Finally, Section~\ref{Conclusion} summarizes our conclusions and outlines the directions for future work.

\begin{figure}
\includegraphics[width=1.0\columnwidth]{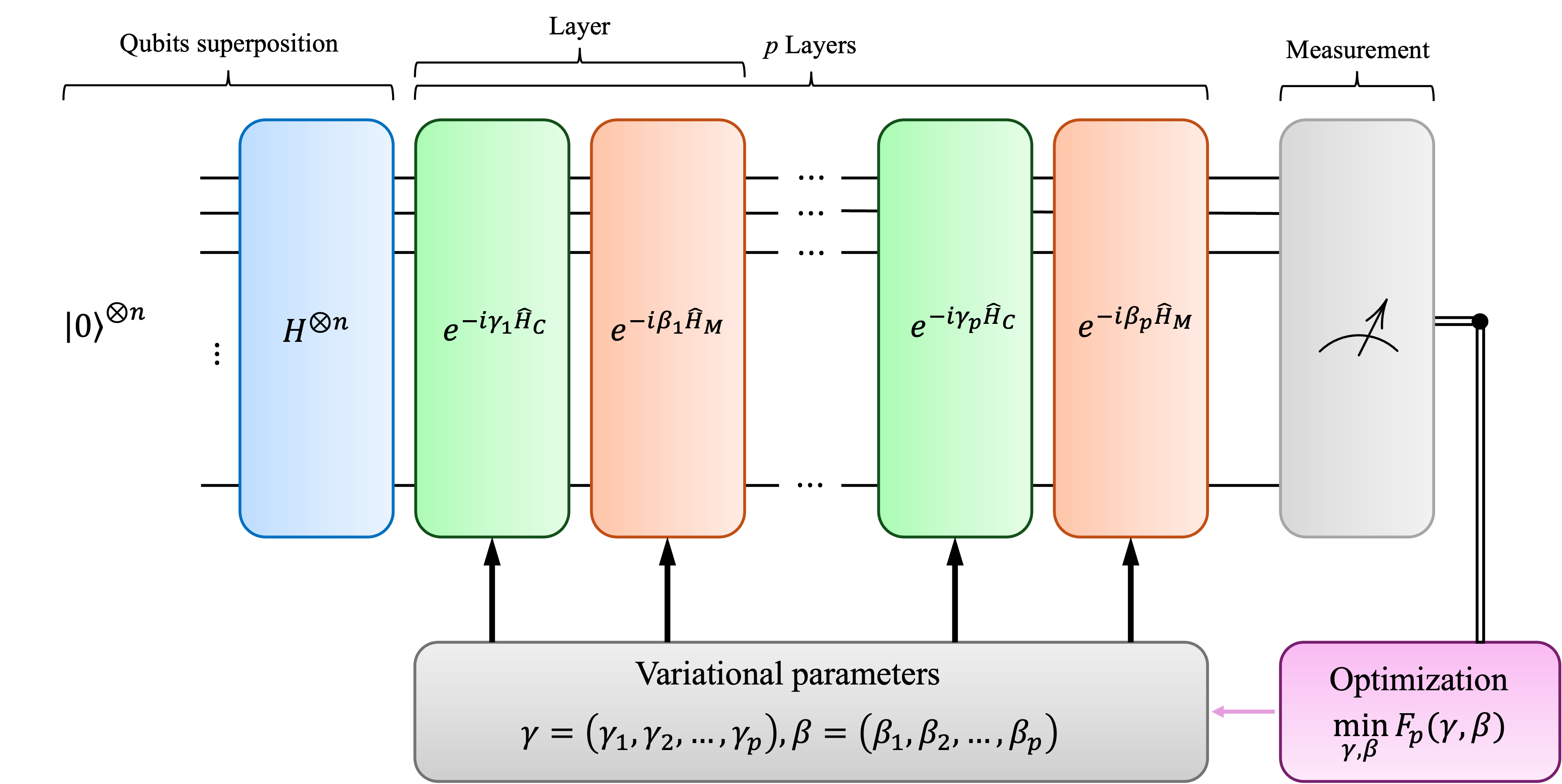}
\caption {\label{Fig1} (Color online) Schematic diagram of the workflow for the Quantum Approximate Optimization Algorithm (QAOA) with $p$ layers. $H_C$ and $H_M$ represent the cost Hamiltonian and mixing Hamiltonian, respectively. The variational parameters $\gamma_k$ and $\beta_k$ at layer index $k$ are optimized using a classical optimizer. The initial spin configuration is prepared in an equal superposition of all possible spin states by applying the Hadamard operator to each state.} 
\end{figure}

\section{Model and Simulation\label{Model}}

The Hamiltonian of the frustrated Ising model on a two-dimensional $L \times L$ square lattice, incorporating competing nearest-neighbor FM interactions and diagonal AF interactions, is given by
\begin{equation}\label{Eq1}
H = -J_1 \sum_{\langle i,j \rangle} \sigma_i^z \sigma_j^z - J_2 \sum_{\langle\langle i,j \rangle\rangle} 
\sigma_i^z \sigma_j^z,
\end{equation}
where the nearest- and diagonal-neighbors are denoted by $<i,j>$ and $<<i,j>>$, respectively. We assume $J_1>0$ and $J_2<0$, corresponding to preferences of FM and AF, respectively. Unlike classic algorithms, which assume each spin variable takes discrete values of $+1$ or $-1$, the Ising Hamiltonian must be reformulated in terms of quantum gates that are implemented in a quantum ansatz, where spins are represented by the simulated qubits~\cite{Jin2012, Jin2013, Park2022, Lee2024, Romer2025}. The spin interactions are then translated into simulated qubit interactions. We reformulated the frustrated Ising model as a combinatorial optimization problem, aiming to find the minimum (or maximum) of a cost function $C(z)$, where $z$ is a classic bitstring. In this formulation, the cost Hamiltonian $H_C$, expressed using Pauli-Z operators, is given by
\begin{equation}\label{eqn:cost_Hamiltonian}
H_C = -J_1 \sum_{\langle i,j \rangle} Z_i Z_j - J_2 \sum_{\langle\langle i,j' \rangle\rangle} Z_i Z_j',
\end{equation}
where $Z_i$ is the Pauli-Z operator acting on the qubit $i$ and $Z_i Z_j$ represents a two-qubit interaction. The corresponding mixing Hamiltonian $H_M$, which introduces quantum fluctuations, is defined as
\begin{equation}\label{eqn:mixing_Hamiltonian}
H_M = \sum_{j} X_j,
\end{equation}
where $X_j$ is the Pauli-X operator acting on the simulated qubit $j$.

\begin{figure}
\includegraphics[width=1.0\columnwidth]{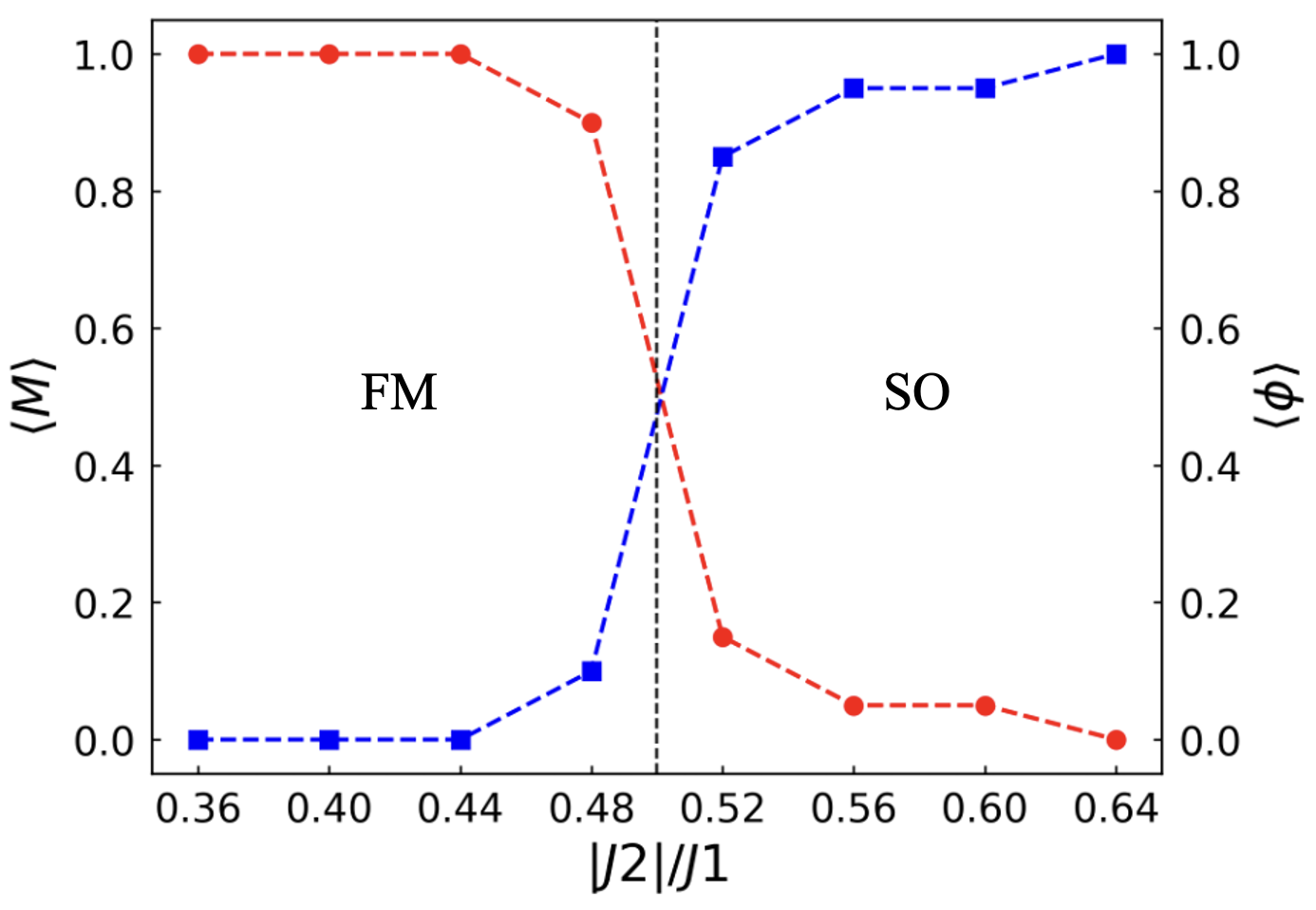}
\caption {\label{Fig2} (Color online) The ferromagnetic (FM) order parameter $\langle M \rangle$ and the stripe antiferromagnetic (AF) order parameter $\langle \phi \rangle$ are plotted as functions of $\vert J_2 \vert/J_1$ for the $4 \times 4$ frustrated Ising model using the QAOA with $p=15$. The FM phase appears below $\vert J_2 \vert/J_1=0.5$, while the stripe AF phase emerges above this value. A phase transition occurs at $\vert J_2 \vert/J_1=0.5$, where the energy gap between the ground state and the first excited state vanishes.}
\end{figure}

Based on this formulation, we constructed a $p$-layer QAOA ansatz as illustrated in Fig.~\ref{Fig1}. Here, $U_C(\gamma_k)$ and $U_M(\beta_k)$ denote unitary operators generated by the cost and mixing Hamiltonians, respectively, at layer index $k$. These alternating unitary layers form a parameterized quantum ansatz, which functions as a time-evolution operator. The resulting quantum state is written as $\vert\gamma_k,\beta_k\rangle$, parameterized by $2p$ variational parameters. The goal is to find optimal $\gamma_k$ and $\beta_k$ that minimize the expectation value of the cost Hamiltonian $\langle \gamma_k, \beta_k \vert H_C \vert \gamma_k, \beta_k \rangle$ A classical optimizer is employed to iteratively update $\gamma_k$ and $\beta_k$ based on measurements obtained from the quantum ansatz.

We initialized the quantum ansatz with an equal superposition of all possible spin configurations by applying Hadamard operator. The initial parameters $\gamma_k$ and $\beta_k$ were set close to zero. To optimize these parameters, we employed the classical Broyden–Fletcher–Goldfarb–Shanno optimizer~\cite{broyden1970, fletcher1970, goldfarb1970, Shanno1970}. The quantum measurements were performed $40$ times to estimate the expectation values.

\section{Results\label{Result}}

\begin{figure}
\includegraphics[width=1.0\columnwidth]{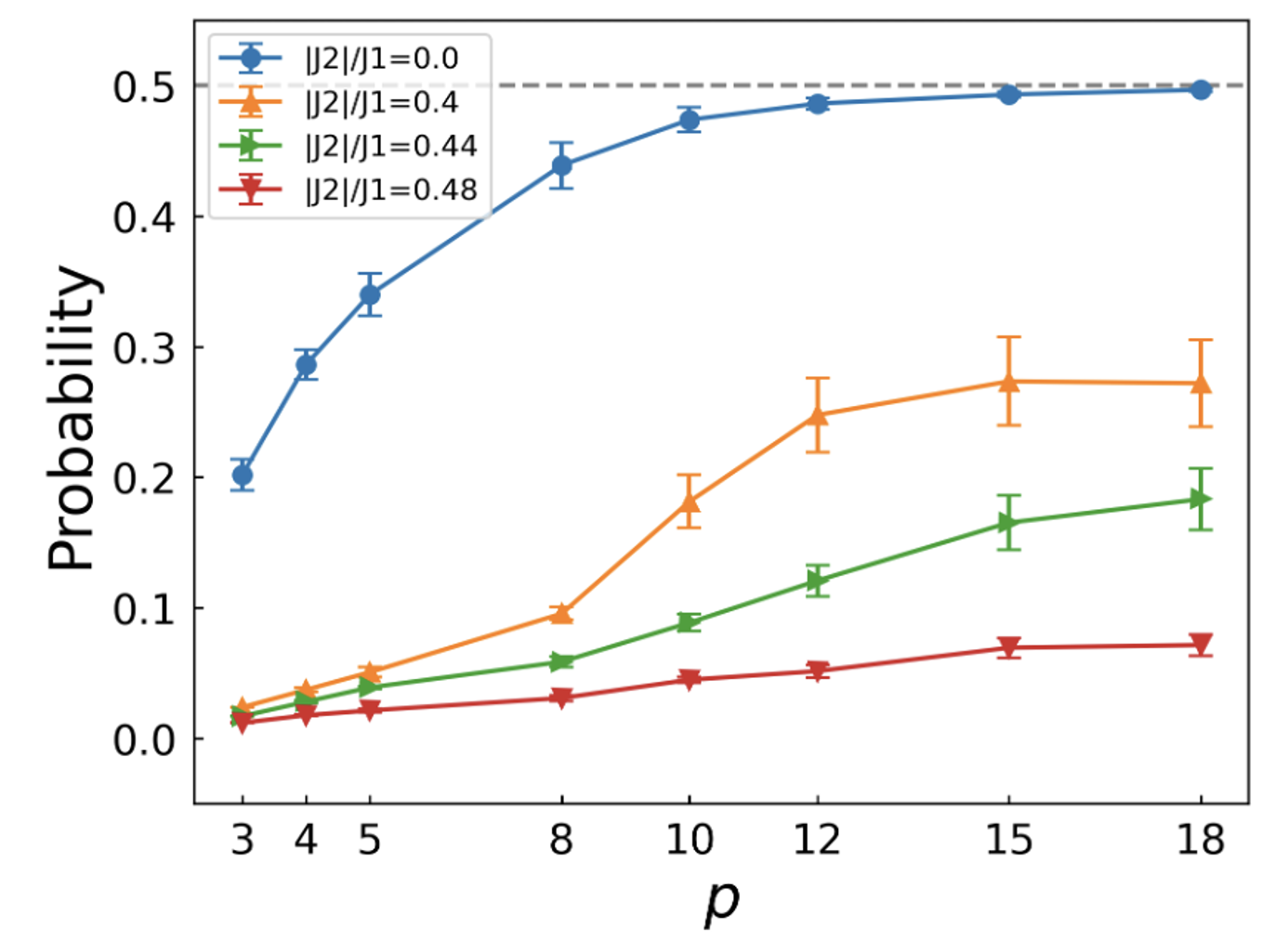}
\caption {\label{Fig3} (Color online) The probability of obtaining the ground state for each QAOA sample set is shown as a function of the number of layers $p$. As $p$ increases, the probability of measuring the ground state also increases. For $\vert J_2 \vert/J_1=0.0$, the probability converges to $0.5$ at $p=18$, reflecting the two degenerate spin configurations that form the ground state. As $\vert J_2 \vert/J_1$ approaches the phase transition point, the converged probabilities at large $p$ remain below $0.5$. The number of sample sets used in each case is $40$.}
\end{figure}

We first computed the FM order parameter $\langle M \rangle$ and the stripe AF order parameter $\langle \phi \rangle$ using QAOA with $p=15$. To evaluate the order parameters, we selected the spin configurations corresponding to the lowest energy obtained from QAOA measurements. Here, the lowest energy in the ensemble does not guarantee the exact ground state energy. Both order parameters as functions of $\vert J_2 \vert/J_1$ are shown in Fig.~\ref{Fig2}. Consistent with previous results, we confirmed that FM states with $\langle M \rangle \approx 1.0$ appear for $\vert J_2 \vert/J_1 < 0.5$, while stripe AF states $\langle \phi \rangle \approx 1.0$ emerge for $\vert J_2 \vert/J_1>0.5$. Strong fluctuations of order parameters are observed near a quantum phase transition at $\vert J_2 \vert/J_1=0.5$.

Next, we calculated the probability of obtaining the lowest energy state as a function of $p$ for several values of $\vert J_2 \vert/J_1$, as shown in Fig.~\ref{Fig3}. At $\vert J_2 \vert/J_1=0.0$, the maximum probability is limited to $0.5$ due to spin symmetry: one ground state configuration with all spins up, and another with all spins down. At this point, even a small value of $p$ (corresponding to the mean field approximation) is sufficient to find the exact lowest energy state with high probability. As $\vert J_2 \vert/J_1$ increases, the probabilities gradually converge to values below $0.5$. We think that this behavior indicates that quantum fluctuations become more significant with increasing $\vert J_2 \vert/J_1$, corresponding to a reduction in $\Delta/J_1$.

\begin{figure}
\includegraphics[width=1.0\columnwidth]{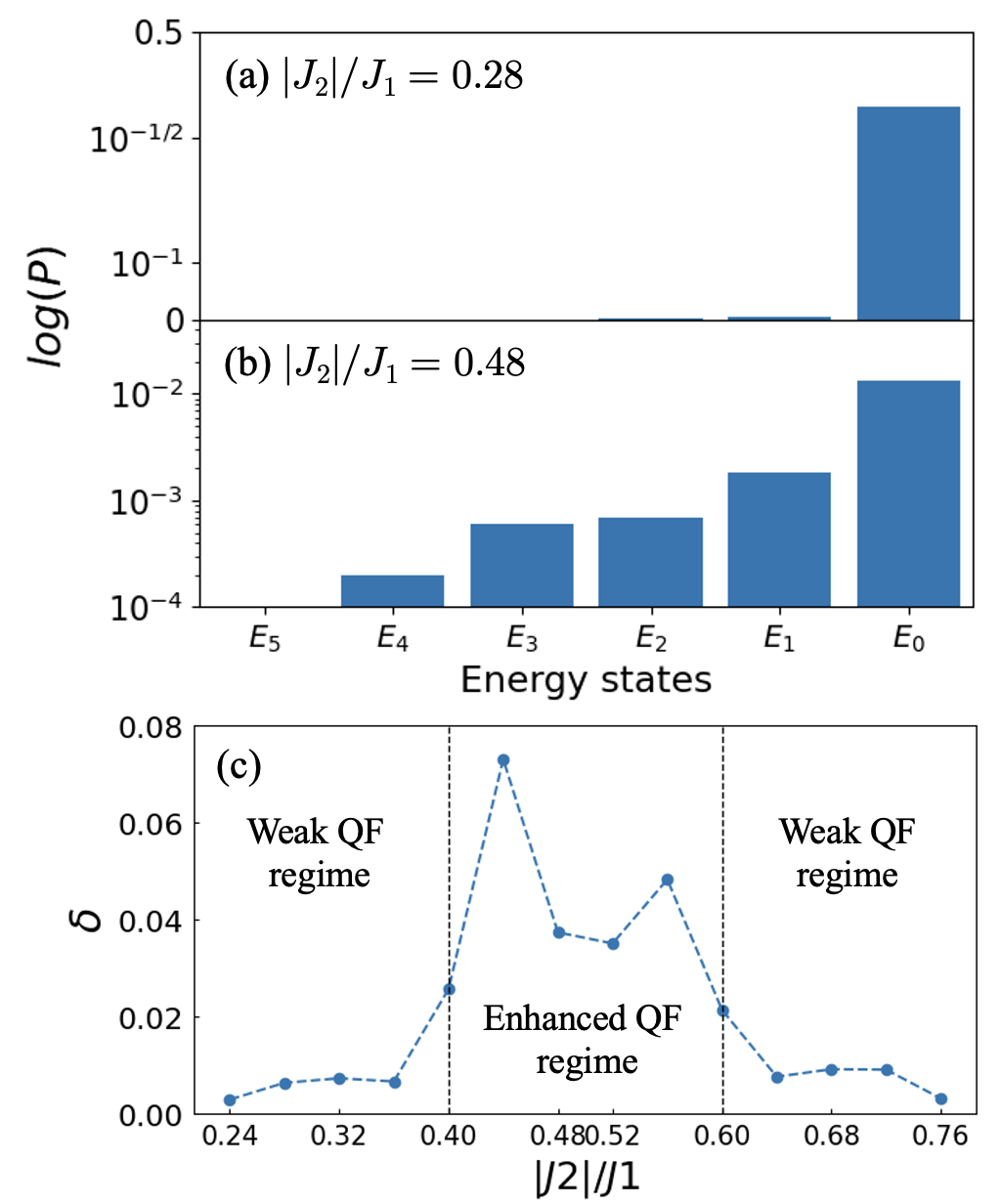}
\caption {\label{Fig4} (Color online) (a) and (b) show the probabilities on a logarithmic scale of several dominant energy states for $\vert J_2 \vert/J_1=0.28$ and $\vert J_2 \vert/J_1=0.48$, respectively. (c) A novel physical quantity $\delta$, defined as $\delta = \langle E \rangle - E_0$, is used to characterize quantum effects. The $\delta$ values are suppressed in the weakly frustrated regime, while they increase significantly in the strongly frustrated regime. 'QF' means quantum fluctuation.}
\end{figure}

To clarify the quantum effects induced by frustration, we plotted the probabilities (on a logarithmic scale) of several dominant energy states for $\vert J_2 \vert/J_1=0.28$ and $\vert J_2 \vert/J_1=0.48$ in Figs.~\ref{Fig4}(a) and (b), respectively. At $J_2/J_1=0.28$ in Fig.~\ref{Fig4}(a), only a single dominant probability is observed, indicating suppressed quantum fluctuations. The corresponding spin configuration with the highest probability is expected to represent the correct FM ground state. In contrast, for $\vert J_2 \vert/J_1=0.48$ in Fig.~\ref{Fig4}(b), multiple dominant probabilities appear due to enhanced quantum fluctuations, and the spin configuration with the highest probability does not necessarily correspond to the true FM ground state. To quantify these quantum effects, we introduced a novel physical quantity $\delta = \langle E \rangle - E_0$, which serves to distinguish between the classical and quantum regimes. Here, $\langle E \rangle$ and $E_0$ represent the ensemble averaged energy and the exact ground state energy, respectively. Fig.~\ref{Fig4}(c) shows $\delta$ as a function of $\vert J_2 \vert/J_1$. 'QF' means quantum fluctuation. The $\delta$ values remain nearly zero for $\vert J_2 \vert/J_1 < 0.36$ in the FM phase and for $\vert J_2 \vert/J_1 > 0.64$ in the stripe AF phase, indicating weak quantum fluctuations in both regimes. In contrast, $\delta$ increases rapidly in the intermediate region $0.36<\vert J_2 \vert/J_1<0.64$, signaling the emergence of significant quantum fluctuations near the phase transition. These results demonstrate that QAOA can successfully capture quantum behaviors that are beyond the reach of classical algorithms.

\section{Conclusion\label{Conclusion}}
We investigated the frustrated Ising model on a two-dimensional $4 \times 4$ square lattice using the QAOA approach. Consistent with previous classical computations, we observed a phase transition between the FM and stripe AF phases at $\vert J_2 \vert/J_1=0.5$. The primary goal of this study is to quantify how frustration contributes to quantum behavior and to examine whether QAOA can capture these effects beyond the capabilities of classical algorithms. We introduced a quantitative measure defined as $\delta = \langle E \rangle - E_0$, which reflects the deviation of the measured energy from the exact ground-state energy, in order to evaluate quantum effects induced by frustration. The parameter $\delta$ is used to distinguish between classical and quantum regimes. We found that quantum behavior is enhanced in $0.36<\vert J_2 \vert/J_1<0.64$, signaling the emergence of significant quantum fluctuations near the phase transition. Finally, we found that quantum measurements under QAOA nearly fail to access the first excited states in weakly frustrated regions. In contrast, near the quantum phase transition point, excited states are predominantly observed, indicating enhanced quantum fluctuations.

As future work, we plan to perform quantum measurements on both a gate-based quantum computer and the D-Wave quantum annealer for the frustrated Ising model, using various values of the parameter $p$ and annealing time. This will allow us to investigate the extent of quantum errors in gate-based quantum hardware by comparing the results with those from classical simulations, which are free from hardware-induced errors. Additionally, we will compare these results with those obtained from the D-Wave quantum annealer, which operates via a continuous quantum adiabatic process, by tuning the annealing time. QAOA measurements on the gate-based quantum computer and quantum annealing measurements on the D-Wave system are each constrained by hardware limitations—namely, a finite number of variational layers in QAOA and a limited annealing time in quantum annealing. Ultimately, this comparison is expected to provide valuable insights into the relationship between quantum circuit depth and annealing time, as well as into the practical accuracy and applicability of both quantum computing platforms.

\section{Acknowledgements\label{Ack}}
This work was supported by Institute of Information and communications Technology Planning Evaluation 
(IITP) grant funded by the Korean government (MSIT) (No. RS-2023-0022952422282052750001) and by the 
National Research Foundation of Korea (NRF) grant funded by the Korea government (MSIT) 
(RS-2024-00453370).


\begin{thebibliography}{99}
\bibitem{Jin2012} S. Jin, A. Sen, and A. W. Sandvik, Ashkin-Teller Criticality and Pseudo-First-Order Behavior in a Frustrated Ising Model on the Square Lattice, Phys. Rev. Lett. {\bf 108}, 045702 (2012).

\bibitem{Jin2013} S. Jin, A. Sen, W. Guo, and A. W. Sandvik, Phase transitions in the frustrated Ising model on the square lattice, Phys. Rev. B {\bf 87}, 144406 (2013).

\bibitem{Park2022} H. Park and H. Lee, Frustrated Ising model on D-wave quantum annealing machine, J. Phys. Soc. Jpn. {\bf 91}, 074001 (2022).

\bibitem{Lee2024} J. Lee, S. Kim, J. Kim, Frustrated Ising model with competing interactions on a square lattice, Phys. Rev. B {\bf 109}, 064422 (2024).

\bibitem{Romer2025} D. Bayo, B. \c civitcio\u{g}lu, J. Webb, A. Honecker, and R. Romer, Machine Learning of Phases and Structures for Model Systems in Physics, J. Phys. Soc. Jpn. {\bf 94}, 031002 (2025).

\bibitem{Preskill2018} J. Preskill, Quantum Computing in the NISQ era and beyond, Quantum {\bf 2}, 79 (2018). 

\bibitem{Bharti2022} K. Bharti, A. Cervera-Lierta, T. H. Kyaw, T. Haug, S. Alperin-Lea, A. Anand, M. Degroote, H. Heimonen, J. S. Kottmann, T. Menke, W.-K. Mok, S. Sim, L.-C. Kwek, and A. Aspuru-Guzik, Noisy intermediate-scale quantum algorithms, Rev. Mod. Phys. {\bf 94}, 015004 (2022).

\bibitem{arute2019} F. Arute, K. Araya, R. Babbush, D. Bacon, J. C. Bardin, R. Barends, R. Biswas, S. Boixo, et al., Quantum supremacy using a programmable superconducting processor, Nature {\bf 574}, 505 (2019).

\bibitem{postler2022} L. Postler, S. Heu{\ss}en, I. Pogorelov, M. Rispler, T. Feldker, M. Meth, C. D. Marciniak, R. Stricker, M. Ringbauer, R. Blatt, P. Schindler, M. M\"{u}ller, and T. Monz, Demonstration of fault-tolerant universal quantum gate operations, Nature {\bf 605}, 675--680 (2022).

\bibitem{nation2025} P. D. Nation, A.A. Saki, S. Brandhofer, L. Bello, S. Garion, M. Treinish, and A. Javadi-Abhari, Benchmarking the performance of quantum computing software for quantum circuit creation, manipulation and compilation, Nat. Comput. Sci. {\bf 5}, 427--435 (2025).

\bibitem{Johnson2011} M. W. Johnson, M. H. S. Amin, S. Gildert, T. Lanting, F. Hamze, N. Dickson, R. Harris, A. J. Berkley, J. Johansson, P. Bunyk, et al., Quantum annealing with manufactured spins, Nature {\bf 473}, 194 (2011).

\bibitem{Higgott2019} O. Higgott, D. Wang, and S. Brierley, Variational Quantum Computation of Excited States, Quantum {\bf 3}, 156 (2019).

\bibitem{King2021} A. D. King, J. Raymond, T. Lanting, et al., Scaling advantage over path-integral Monte Carlo in quantum simulation of geometrically frustrated magnets, Nat. Commun. {\bf 12}, 1113 (2021).

\bibitem{Park2024(1)} H. Park, and H. Lee, Determination of Optimal Chain Coupling made by Embedding in D-Wave Quantum Annealer, AVS Quantum Sci. {\bf 6}, 033804 (2024).

\bibitem{Park2024(2)} H. Park, and H. Lee, Quantum eigensolver on extension of optimized binary 
configurations, {\bf 110}, 205142 (2024).


\bibitem{farhi2014} E. Farhi, J. Goldstone and S. Gutmann, A quantum approximate optimization algorithm, arXiv:1411.4028 (2014).

\bibitem{farhi2015} E. Farhi, J. Goldstone and S. Gutmann, A Quantum Approximate Optimization Algorithm Applied to a Bounded Occurrence Constraint Problem, arXiv:1412.6062 (2015).

\bibitem{farhi2019} E. Farhi, A. W. Harrow, Quantum Supremacy through the Quantum Approximate Optimization Algorithm, arXiv:1602.07674 (2019).

\bibitem{choi2019} J. Choi, and J. Kim, A tutorial on quantum approximate optimization algorithm (QAOA): Fundamentals and applications, 2019 international conference on information and communication technology convergence (ICTC), IEEE, 138--142 (2019).

\bibitem{farhi2020} E. Farhi, D. Gamarnik, and S. Gutmann, The Quantum Approximate Optimization Algorithm Needs to See the Whole Graph: Worst Case Examples, arXiv:2005.08747 (2020).

\bibitem{zhou2020} L. Zhou, S.-T. Wang, S. Choi, H. Pichler, and M. Lukin, Quantum approximate optimization algorithm: Performance, mechanism, and implementation on near-term devices, Phys. Rev. X {\bf 10}, 021067 (2020).

\bibitem{moussa2022} C. Moussa, H. Wang, T. B\"{a}ck, and V. Dunjko, Unsupervised strategies for identifying optimal parameters in quantum approximate optimization algorithm, EPJ Quantum Technol. {\bf 9}, 11 (2022).

\bibitem{blekos2024} K. Blekos, D. Brand, A. Ceschini, C.-H. Chou, R.-H. Li, K. Pandya, and A. Summer, A review on quantum approximate optimization algorithm and its variants, Phys. Rep. {\bf 1068}, 1--66 (2024).

\bibitem{farhi2000} E. Farhi, J. Goldstone, S. Gutmann, and M. Sipser, Quantum computation by adiabatic evolution, arXiv:quant-ph/0001106 (2000).

\bibitem{farhi2001} E. Farhi, J. Goldstone, S. Gutmann, J. Lapan, A. Lundgren, and D. Preda, A quantum adiabatic evolution algorithm applied to random instances of an NP-complete problem, Science {\bf 292}, 472--475 (2001).












\bibitem{broyden1970} C. G. Broyden, The Convergence of a Class of Double-rank Minimization Algorithms 1. General Considerations, IMA J. Appl. Math. {\bf 6}, 76--90 (1970).

\bibitem{fletcher1970} R. Fletcher, A new approach to variable metric algorithms, Comput. J. {\bf 13}, 317--322 (1970).

\bibitem{goldfarb1970} D. Goldfarb, A family of variable-metric methods derived by variational means, Math. Comp. {\bf 24}, 23--26 (1970).

\bibitem{Shanno1970} D. F. Shanno, Conditioning of quasi-Newton methods for function minimization, Math. Comp. {\bf 24}, 647--656 (1970).

\end{thebibliography}
\end{document}